\documentclass{emulateapj}
\usepackage{natbib}
\usepackage{color}

\definecolor{purple}{rgb}{0.5, 0.0, 0.5}

\begin{document}

%\title{The Role of Magnetic Complexity in Stellar Spin Down}
%\title{The Role of Magnetic Complexity in the Spin Down of Young Stars}
\title{The Revolution Revolution$^\dagger$: magnetic morphology driven spin-down}

\author{ C. Garraffo\altaffilmark{1}, J.~J. Drake\altaffilmark{1}, A. Dotter\altaffilmark{1}, J. Choi \altaffilmark{1},  D.~J. Burke\altaffilmark{1}, S.~P. Moschou\altaffilmark{1}, J.~D. Alvarado-G\'omez\altaffilmark{1}, V.~L.\ Kashyap\altaffilmark{1}, and O. Cohen\altaffilmark{1,2}}

\altaffiltext{1}{Harvard-Smithsonian Center for Astrophysics, 60 Garden St. Cambridge, MA 02138, USA}
\altaffiltext{2}{Lowell Center for Space Science and Technology, University of Massachusetts Lowell, 600 Suffolk Street, Lowell, MA 01854, USA}
%%%%%%%%%%%%%%%%%%%%%%%%%%%%%%%%%%%%%%%%%%%%%%%%%%%%%%%%%%%%%%%%%%%%%%%%%%%%%%%%%%%%%
% Abstract
%%%%%%%%%%%%%%%%%%%%%%%%%%%%%%%%%%%%%%%%%%%%%%%%%%%%%%%%%%%%%%%%%%%%%%%%%%%%%%%%%%%%%

\begin{abstract}
Observations of young open clusters show a bimodal distribution of rotation periods that has been difficult to explain with existing stellar spin-down models. 
Detailed MHD stellar wind simulations have demonstrated that surface magnetic field morphology has a strong influence on wind-driven angular momentum loss. Observations suggest that faster rotating stars store a larger fraction of their magnetic flux in higher-order multipolar components of the magnetic field.  In this work, we present an entirely predictive new model for stellar spin-down that accounts for the stellar surface magnetic field configuration. We show how a magnetic complexity that evolves from complex toward simple configurations as a star spins down can explain the salient features of stellar rotation evolution, including the bimodal distribution of both slow and fast rotators seen in young open clusters.  

\end{abstract}

\keywords{stars: rotation --- stars: magnetic field --- stars: evolution }

%@arxiver{singlestar_new_S.pdf, singlestar_new.pdf,M37_346.pdf,Hyades_500.pdf,Praesepe_550.pdf}

%%%%%%%%%%%%%%%%%%%%%%%%%%%%%%%%%%%%%%%%%%%%%%%%%%%%%%%%%%%%%%%%%%%%%%%%%%%%%%%%%%%%%
% Introduction
%%%%%%%%%%%%%%%%%%%%%%%%%%%%%%%%%%%%%%%%%%%%%%%%%%%%%%%%%%%%%%%%%%%%%%%%%%%%%%%%%%%%%
\section{INTRODUCTION}
\label{sec:Intro}
%\begin{onecolumn} %JJD: commented this out as it induced compiled errors
\let\thefootnote\relax\footnote{$^\dagger$Title inspired by  C.~L.~Davis PhD Thesis \textit{Revolution evolution: tracing angular momentum during star and planetary system formation}, St. Andrews University.}
%\end{onecolumn}
The rotation evolution of stars has been extensively studied over the last five decades but remains one of the most challenging and open problems in stellar astrophysics. Rotation is relevant for stellar evolution itself, for stellar age determination via the technique of ``gyrochronology'' \citep{Skumanich:72, Soderblom:83, Barnes:03, Meibom.etal:15}, and is the driver of stellar magnetic activity. Growing realisation that the latter plays a crucial role in exoplanet detection \citep[see, for example,][]{Hatzes:13a, Hatzes:13b, Hatzes:16, Donati.etal:16} and on exoplanet atmospheric evolution and habitability \citep{Sanz-Forcada.etal:11, Lammer.etal:12,Chadney.etal:16, Garraffo.etal:16b, Garraffo.etal:17,Cohen.etal:18, Chadney.etal:17}, combined with the increasing number of precise measurements of rotation periods of stars, presents a renewed motivation to revise our understanding of stellar rotation. 

Magnetic braking is the dominant mechanism by which Sun-like and later type stars spin down, and it is determined by the magnetic fields on their surfaces \citep{Weber.Davis:67, Kawaler:88}. Stellar rotation fuels magnetic activity through dynamo action and, in turn, activity controls spin-down rates.  This self-regulating mechanism results in a relationship between rotation period and mass that evolves with time.
Observations of open clusters (OCs) of known ages indeed show stars for which rotation angular velocities ($\Omega$) follow the Skumanich spin down law, $\Omega \sim t^{-1/2}$, but they also show persistent fast rotators whose origin remains a mystery.  
%While we can reproduce the I-branch reasonably well with semi-empirical models \cite{Weber.Davis:67, Kawaler:88, Barnes:03, Barnes:10, Barnes.Kim:10, Matt.etal:12}, what remains a mystery is the fast rotators branch that gets less populated with time, and the gap between the two branches that seems to be a fast transition between them. 

%The observed shape of the Skumanich-like branch suggests a stellar mass dependence of the spin-down torque that is best captured by the convective turnover timescale. This was supported by theoretical work by \cite{Durney.Latour:78} and \cite{Noyes.etal:84}.  The usual assumption is that, in this branch, rotation periods can be described by the product of a color dependent function $f(B-V)$, related to the convective turnover time, and $t^{-1/2}$, representing the Skumanich spin-down law. This is a simplification that has been shown to have shortcomings but still provides a reasonable fit for the observed clusters.  
%Much less is known about the fast-rotators branch. The large spread on rotation periods of young stars suggests they have a range of wind properties that disappears with stellar age. 
The implication of the observations is that some stars undergo a fairly rapid spin-down whereas others of the same mass and age do not.  A considerable amount of recent theoretical work has been aimed at explaining this OC rotational bimodality. 
One of the first
empirical models is The Double Zone Model \citep[see][and references therein]{Spada.etal:11} and recent variations of it \citep{Reiners.Mohanty:12, Gallet.Bouvier:13}.  This model is based on the analytical prescription for the stellar wind torque given by \cite{Kawaler:88} and includes a bifurcation at a certain critical stellar rotation frequency $\Omega_{crit}$. %but does not provide a physical mechanism for it.  
%Another salient example is the 

The Symmetrical Empirical Model of \citet{Barnes:10} and \citet{Barnes.Kim:10} also uses a bifurcated prescription for the torque, and is based on the idea of a sudden coupling of the stellar radiative core with its convective envelope that results in a fast and dramatic spin-down due to the sudden change in moment of inertia.  The latter has been the standard solution until recently, when the same bimodal behavior was observed for fully convective stars \citep{Newton.etal:16, Douglas.etal:16, Douglas.etal:17}, suggesting that the radiative core does not play a significant role in the sudden change of angular velocity. 
%The most successful models so far are more descriptive than explanatory and fall short of reproducing the morphology of the open-cluster observations.

Recent studies have aimed for a more unified scenario.  While great progress has been made with sophisticated, physics-based, and realistic models \citep{Matt.etal:12, Matt.etal:15, Ardestani.etal:17, Pantolmos.Matt:17}, these still do not recover the bimodal morphology of the color-period diagram. \citet{Johnstone.etal:15b} successfully modeled the {\em spread} in rotation rates but only partially recovered the bimodality of their distributions as reflection of a bifurcated wind torque formula.  \cite{Qureshi.etal:18} recently did a proof-of-concept calculation that shows planet consumption by a star can lead to faster rotation.

\cite{Brown:14} presented the first prescription that successfully reproduces the simultaneous presence of rapid and slow branches, and the sparsely populated gap between them, and matches observations reasonably well. This ``Metastable Dynamo Model" (MDM), is based on the idea that rotating stars fall into two different regimes, one in which the dynamo is strongly coupled to the wind, that accounts for the Skumanich branch, and the other one in which it is weakly coupled and gives rise to the branch of fast rotators. MDM requires a spontaneous and random change of mode from the former to the latter, with a mass-dependent transition probability, and predicts that the angular momentum loss (AML) efficiency in the strongly coupled regime is at least two orders of magnitude larger than that in the weakly coupled regime.  However, this model lacks a physical basis for each regime and a mechanism for the transition between them.   

All existing models for stellar spin-down have, to a large extent, neglected the geometry of stellar surface magnetic fields. An increasing number of Zeeman-Doppler-Imaging (ZDI) observations indicate that young, active stars store a larger fraction of their magnetic flux in higher-order multipole components of the magnetic field, i.e.\ complex field configurations \citep[e. g.][]{Donati:03, Donati.Landstreet:09, Marsden.etal:11, Waite.etal:15, Alvarado-Gomez.etal:15}. 
What effects this might have on spin-down rates is, then, of considerable interest and has been the subject of a handful of recent studies \citep{Reville.etal:15a, Garraffo.etal:15, Garraffo.etal:16a}. 

\cite{Garraffo.etal:15} have shown that the complexity of the large scale magnetic field can dramatically reduce the angular momentum loss rates by a few orders of magnitude.  As a consequence, one should expect their magnetic braking efficiency to be lower than that of their slower rotating relatives. \cite{Garraffo.etal:15} pointed out that this complexity provides a physical basis for the MDM model and, therefore, might naturally explain the bimodal distribution of rotation periods observed in OCs. \cite{Garraffo.etal:16a} (from here on, CG16) quantified this effect, providing scaling laws for stellar angular momentum loss rates as a function of complexity together with a prescription for applying them to real stars. Later, \cite{Finley.Matt:17, Finley.Matt:18, See.etal:18} used scaling laws based on a set of thermally driven polytropic wind simulations to explore the effect of mixed modes. 
%Our model includes the additional heating and momentum terms that are responsible for the observed solar corona heating and solar wind acceleration.  These differences reflect on a quantitative disagreement, however, the qualitative conclusions of both works, finding that the dominant order is the lowest one, are consistent.

Recently, \cite{Vansaders.etal:16} reported an additional later deviation from standard Gyrochronology.  They find that after stars reach Rossby number $\sim 2$ ($Ro = P/\tau$, where $\tau$ is convective turnover time), they rotate faster than expected from the Skumanich law and, therefore, appear to be losing angular momentum less efficiently. This, together with a reported deficit of observed rotation periods longer than the Sun among solar type stars \citep[see][and reference therein]{Vansaders.etal:18}, suggests that a very efficient magnetic braking suppression takes place at later ages.   This could be due to a sudden decrease in magnetic field strength but it would require a very dramatic change in the dynamo.  Instead, it can be explained by a smooth increase in magnetic field complexity at late times that, given the steep dependency of angular momentum loss efficiency with complexity, would result in a sharp magnetic braking suppression.

% \textbf{ While this could be due to a sudden decrease in magnetic field strength, it would require a very dramatic change in the dynamo.  This sharp magnetic braking interruption can be explained, instead, by a smooth increase in magnetic field complexity at late times.  As discussed before, this would results in a strong suppression of angular momentum loss efficiency. }

In this work, we present a new prescription for the rotation evolution of young, active stars that includes the modulation of spin-down rates derived by CG16.
We confirm that accounting for magnetic complexity in our spin-down models results in the bimodal distribution observed in OCs and its observed evolution over time.

This paper is organized as follows. In Section~\ref{sec:Model} we describe the spin-down model, in Section~\ref{sec:Methods} we detail the method we used to generate population synthesis, in Section~\ref{sec:OCO} we describe the observations against which we will compare our populations, and in Sections~\ref{sec:Results} and \ref{sec:Discussion} we discuss our results and their implications. Lastly, in Section~\ref{sec:Conclusions}, we summarize our conclusions.

\section{THE MODEL}
\label{sec:Model}

%This naturally results in the three components observed in young open cluster rotation periods: the Skumanich-like branch, the fast rotators branch, and the gap in between;  as well as the right time-evolution of all three. 

%\subsection{MAGNETIC BRAKING PRESCRIPTION}
%\label{sec:mb}

\cite{Weber.Davis:67} provided the first prescription for calculating angular momentum loss rates. Their model assumed spherical symmetry and was
later generalized by simple scaling relations \citep{Kawaler:88, Krishnamurthi.etal:97}. 
%The need of a mass dependent $\Omega_{crit}$ suggests that this prescription fails to capture the mass dependence of the slower rotating stars. (see Barnes.Kim 2010) 
Observations support the idea that the slow rotators branch follows the Skumanich spin-down law, $P \sim t^{1/2}$, where $P$ is the rotation period of the star, and, therefore, rotation periods can be determined as a function of stellar mass and age. 
  
Our prescription for angular momentum loss rates is based on two assumptions. The first assumption is that stars with a dominant dipolar component of the magnetic field follow a Skumanich spin-down law with a mass dependence that reflects the convective turnover time $\tau$.  The latter is currently a standard assumption, based on theoretical work by \cite{Durney.Latour:78}, for all stars regardless of their magnetic field configuration \citep[See, for example,][]{Barnes:03,Barnes.Kim:10,Brown:14}.  The second assumption is that the magnetic fields of fast rotating stars have more complex large-scale geometries, which, as discussed before, is supported by ZDI observations. Magnetic complexity is expected to reduce the angular momentum loss rates significantly \citep{Reville.etal:15a, Garraffo.etal:15} and we employ the scaling laws derived by CG16.  The model presented here then uses a simple magnetic braking prescription based on Skumanich spin-down law together with this magnetic complexity modulation.

The angular momentum loss can be written as
\begin{equation}
\dot{J} = \dot{J}_{Dip} Q_J(n),
\label{eq:J}
\end{equation}
where $\dot{J}_{Dip} $ represents the dipolar losses and $Q_J(n)$ is a modulating factor that accounts for the complexity of the magnetic fields on the stellar surface,  parametrized by $n$ (see CG16 for details).
%where $\dot{J}_{Dip} = \Omega^4 \tau$ represents the dipolar losses and corresponds to the usual prescription for angular momentum loss rates on the Skumanich-like branch, and $Q_J(n)$ represents the magnetic complexity dependence.
%However, in order to reproduce the slow rotators branch, we find that the time dependence is closer to $\Omega \sim t^{0.25}$, resulting in a $\dot{J}_{Dip} \sim \Omega^5 f(B-V)$. Dimensionally, that means that the mass dependence function should be $f(B-V) \sim \tau^3$. 
The dipolar branch, which corresponds to $Q_J=1$, evolves in time following a Skumanich law, $P_{Dip} \sim t^{0.5}$, that translates into the angular momentum loss rate as,
\begin{equation}
 \dot{J}_{Dip} = c \cdot \Omega^3 \tau, 
 \label{eq:J2}
 \end{equation}
 where $c$ is a normalization factor related to the wind efficiency for a dipole, and is well-constrained by observations and stellar spin-down time-scales.  The shape of this branch reflects the color dependence of the convective turnover time $\tau$.
%We have tried a Skumanich type evolution (which is $\Omega^3$ in Jdot) for both a function f(B-V) as in \cite{Mamajek.Hillenbrand:08}, and for just $\tau$ instead of the function. For $\tau$ from tevol tables (which give chuncky pieces because of them being every 0.1 mass) and from Wright et al, we get that tail going up for late type stars in older clusters. 
%$Jdot = \Omega^4 \tau^2$ with $\tau$ from $f(B-V)$ and color being recalculated for each time as a function of Teff (which I extrapolate linearly for lower mass stars that we have no tables), if fits better for younger clusters. If we take $\sqrt{f(B-V)}$  instead, then it fits ok for older clusters but not so much for younger.
%We take $\tau$ from the \textit{Tevol} tables and fit $\tau = \tau(B-V)$ to get a smooth function of color.  
%The following function 
%$(10*(5*\sqrt(abs(BV-0.45))+0.4*(BV-0.45)+3*(BV-0.45)**4))^0.75$ (similar to the one derived by \cite{Mamajek.Hillenbrand:08}) is a good fit for $\tau$ and reproduces reasonably well the I-branch for all open cluster observations if $Jdot \sim \Omega^5$.... for $a=0.15$ and $b=1$ (see Figure~\ref{fig:BV}). 
%\begin{equation}
%\tau \sim f(B-V) =   0.407[(B - V)_0 - 0.495]^{0.325}
%\end{equation}
We use the magnetic complexity modulation factor derived by CG16,
\begin{equation}
Q_J(n) = 4.05 \, e^{-1.4 n}+(n-1)/(60 B \cdot n),
\end{equation}
where $n$ is the complexity of the magnetic field ($n=1$ represents a dipole and is larger for higher complexity) and $B$ represents the magnetic field strength.  The second term becomes important only for $n>7$ at which the spin down rate reaches a plateau (see Figure~3 from CG16). We neglect this effect by imposing $n=7$ as the maximum complexity and, therefore, the above equation simplifies to:
\begin{equation}
Q_J(n) = 4.05 \, e^{-1.4 n}
\label{eq:Q}
\end{equation}

\begin{figure}{}
%%\vspace{2in}
\center
\includegraphics[trim = .5in 2.2in  0.4in 2.5in,clip, width = 0.5\textwidth]{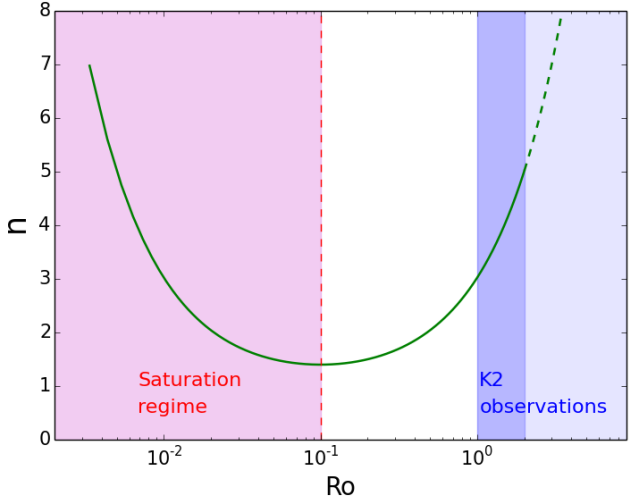}
\caption{Proposed complexity of the magnetic field as a function of Rossby number consistent with Open Cluster and K2 observations. The pink shaded area represents the saturated regime observed in X-rays and the blue shaded area represents the Rossby numbers at which the later deviation from Gyrochronology is observed.}
\label{fig:n}
\end{figure}

As discussed earlier, according to ZDI observations, young, fast rotating stars seem to be more complex and, therefore, we expect $n$ to decrease with rotation period or its dimensionless relative Rossby number, $Ro = P_{rot}/\tau$.  On the other hand, recent {\it Kepler} observations show a deviation from Gyrochronology at $Ro \sim 1 - 2 $, that in this scenario corresponds to a new increase of complexity.  
%In addition, stars with mid Rossby numbers ($0.1< Ro <1-2$)  may present magnetic cycles as a consequence of differential rotation ??. This should result, as is the case for the Sun, in a small oscillatory behavior of its large-scale complexity, with a period of days {\bf ??days? ??} to years.  Short-term behavior does not affect our model and will be erased on evolutionary time-scales. However, the average complexity will be raised if these cycles occur when the complexity is too close to the minimum allowed value $n=1$ that corresponds to a dipole. We reflect this by a slight increase in the function $n(Ro)$ in the regime where differential rotation and, therefore, cycles, are expected: $0.1<Ro < 2$.
We propose a simple function for the complexity of the magnetic field with Rossby number (see Figure~\ref{fig:n}) that reflects the trends suggested by these observations:
\begin{equation}
n= \frac{a}{Ro}+1+bRo,
\label{eq:n}
\end{equation}
where $a = 0.02$ and $b = 2$ are the free parameters in our model and were determined using OCs observations.  The first term represents the decrease in complexity with rotation period for young, fast rotating stars, suggested by ZDI observations.  The constant $1$ is just reflecting the fact that the minimum possible complexity is a pure dipole and has been defined as $n=1$ by CG16. The minimum of this function has been taken to be slightly higher than one given that even the stars on the slow rotators branch, expected to follow Skumanich, are probably not perfect dipoles.  The third term is included to represent the increase in complexity at later times that would explain {\it Kepler} observations and it only becomes important at larger Rossby numbers than the ones most relevant for this work. We emphasize that this term is not necessary for the success of our model in reproducing the young stellar cluster observations shown here.  It will, however, make a difference when modeling older cluster rotation period distributions. 
{}

\begin{figure*}
%%\vspace{2in}
\center
\includegraphics[trim = .3in 2.2in  0.3in 0in,clip, width = 0.45\textwidth]{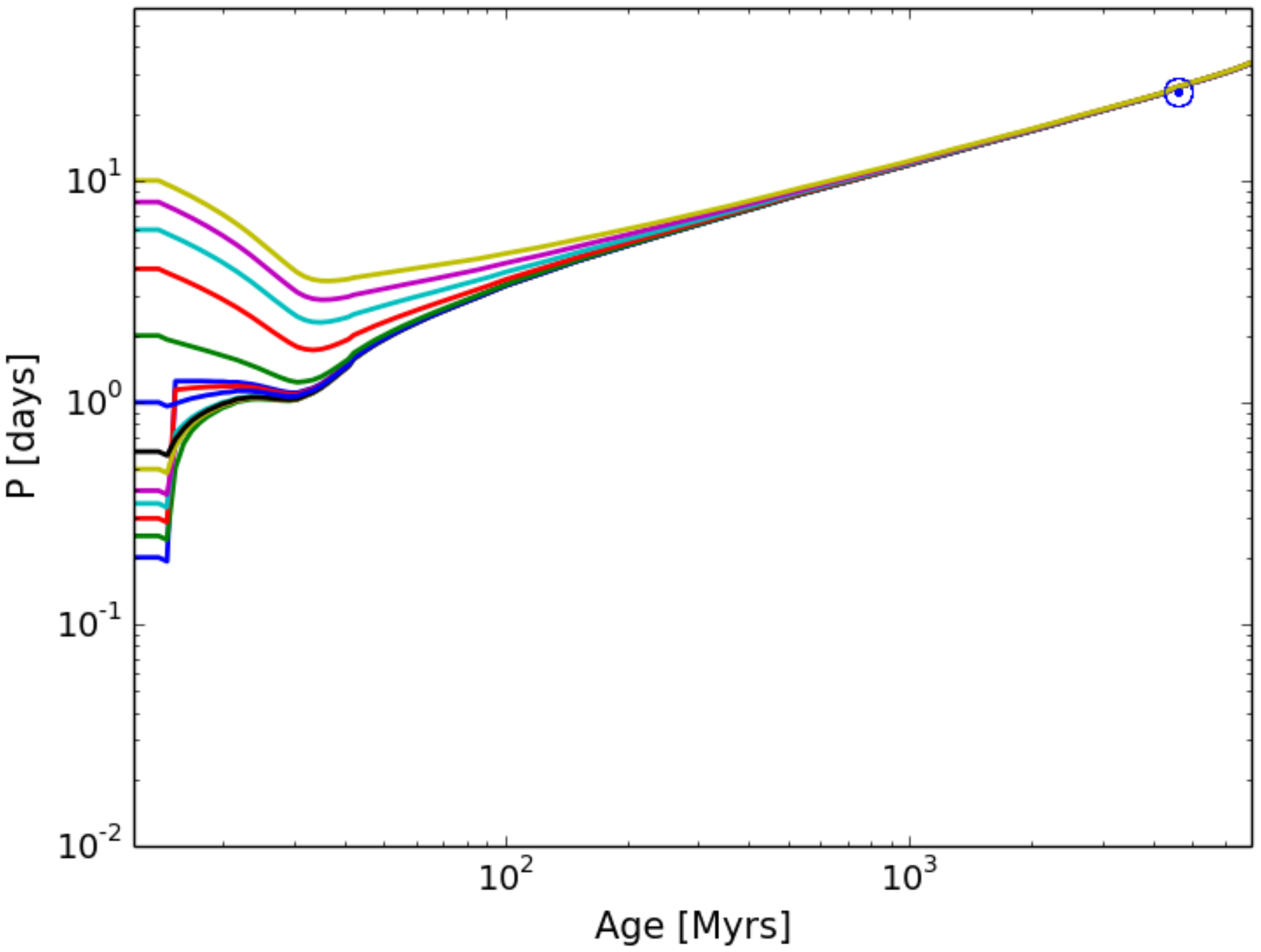}
\includegraphics[trim = .3in 2.2in  0.3in 0in,clip, width = 0.45\textwidth]{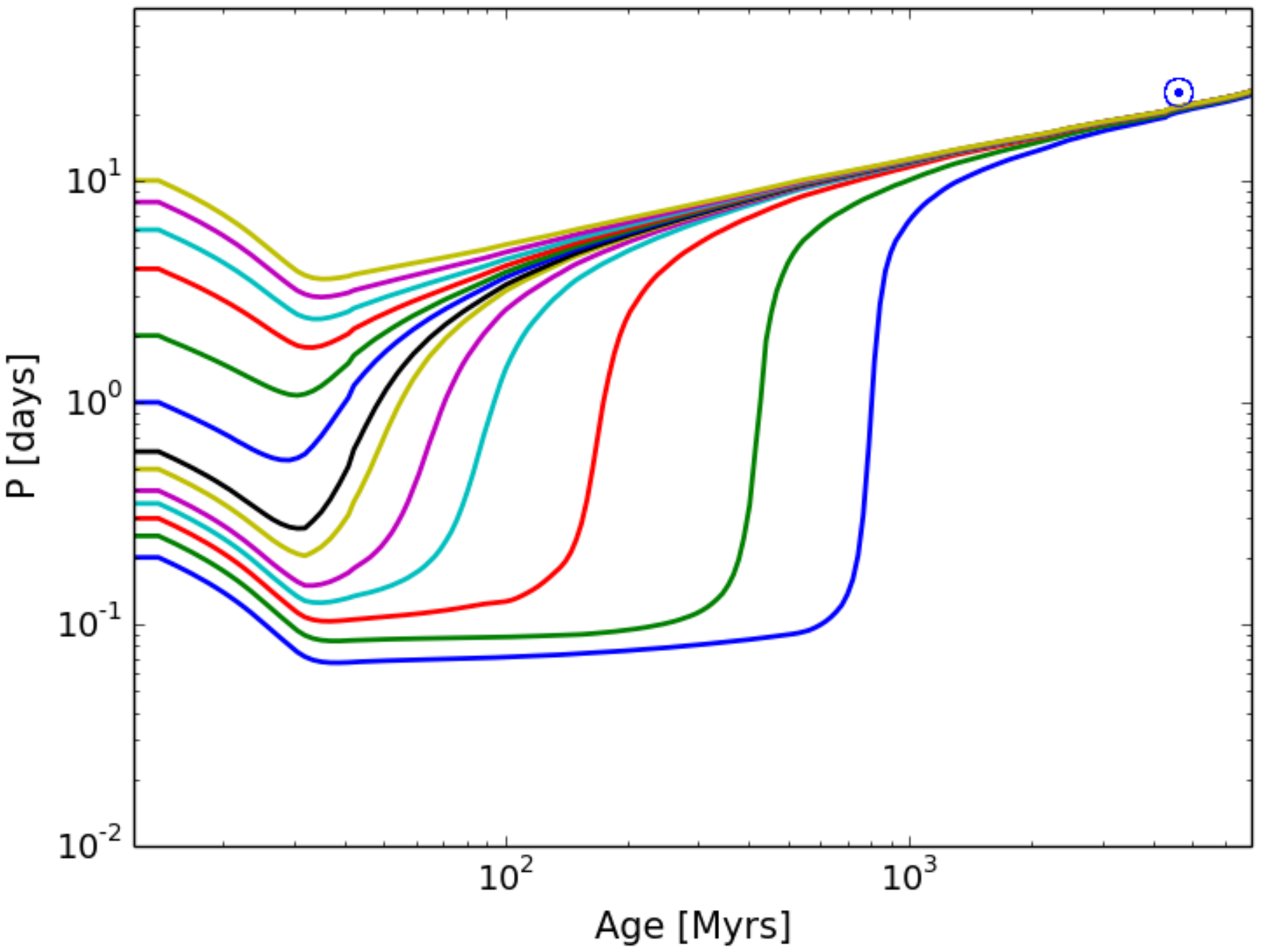}
\caption{Rotation period evolution of a solar mass star for different initial periods. The left panel shows evolution without taking into account the complexity modulation (Skumanich-like spin-down), and the right panel shows the same but including the complexity modulation, which corresponds to our model predictions. In both panels, the present-day Sun is indicated. }
\label{fig:sun}
\end{figure*}

%????? The tau used for this Ro is the Tau from MH08, so it assumes that the convective turnover time depends on color. We could use Tau as a function of mass, as in Wright et al 2011, but that would mean the size of the convective zone doesn't change with time... which might not be true when starting from such early times. 
%Probably, the most accurate treatment would be to use the one fro TEVOL tables but for that we would need a fitted function since we only have Tau for intervals of 0.1 solar masses. This would mean doing it for each time, and it is challenging because tables for different masses have different time steps. ????????
Under this description, the angular momentum loss rate for any given star is a function of just its Rossby number.  In Figure~\ref{fig:sun}, we have reproduced the rotation evolution of a solar mass star starting at 13~Myrs (immediate post-disk phase) for different initial conditions with 
%(right panel, reflecting our predictions) 
and without 
%(left panel) 
considering the magnetic modulation of the angular momentum loss efficiency. 

%in the angular momentum loss equation. 

%If $f(B-V) = \frac{\tau}{Ro}$  then $\dot{J}_{Dip} = \Omega^3 \tau^2 \Omega= \Omega^4 \tau^2$. Now, we have fitted $\tau$ to be $f(B-V)$ but it could also be $f(B-V)^2$. I am trying these things.  In the literature it is not clear that there is a goof fit to the I-branch. \cite{Brown:14} finds that the exponent of f varies.

%We have tried a skumanich type evolution (which is $\Omega^3$ in Jdot) for both a function f(B-V) as in \cite{Mamajek.Hillenbrand:08}, and for just $\tau$ instead of the function. For $\tau$ from tevol tables (which give chuncky pieces because of them being every 0.1 mass) and from Wright et al, we get that tail going up for late type stars in older clusters. 
%$Jdot = \Omega^4 \tau^2$ with $\tau$ from $f(B-V)$ and color being recalculated for each time as a function of Teff (which I extrapolate linearly for lower mass stars that we have no tables), if fits better for younger clusters. If we take $\sqrt{f(B-V)}$  instead, then it fits ok for older clusters but not so much for younger.

%For our fit of Tau $(10*(5*\sqrt(abs(BV-0.45))+0.4*(BV-0.45)+3*(BV-0.45)**4))^0.75$ we get a decent fit if $Jdot \sim \Omega^5$.... for $a=0.15$ and $b=1$ (see Figure~\ref{fig:BV}. 

%\subsection{Disk Locking}
%\label{sec:dl}

\section{POPULATION SYNTHESIS}
\label{sec:Methods}
 
\begin{figure}{}
%%\vspace{2in}
\center
\includegraphics[trim = .3in 2.2in  0.3in 2.5in,clip, width = 0.5\textwidth]{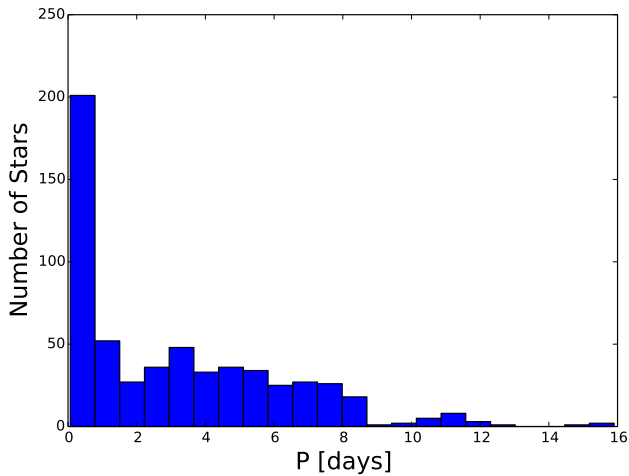}
\caption{Histogram of initial periods from h Persei Cluster.}
%\caption{Histogram of initial periods from Orion Nebula Cluster.}
\label{fig:histogram}
\end{figure}

To compare our model with OC observations,  we generate a population of stars with initial rotation periods and stellar masses from the h Persei Cluster  \citep{Moraux.etal:13}; see Figure~\ref{fig:histogram}).
%a uniform distribution of masses between 0.3 and 1.6 $M_\sun$ and initial rotation period, independent of stellar mass,  
%drawn from Orion Nebula Cluster observations (\citealt{Mermilliod:96}; see Figure~\ref{fig:histogram}) as in \cite{Brown:14}.  

{}

We use Monte Carlo simulations with a time step of $10^4$ years and evolve the population for 1~Gyr.  
We include the evolution of each star's mass, radius, moment of inertia, and effective temperature using the \textit{MIST} evolutionary tracks \citep{Dotter:16, Paxton.etal:11, Paxton.etal:13, Paxton.etal:15}. We do not use the rotation periods from the \textit{MIST} tables, instead, the rotation periods are self-consistently evolved according to our model.  At each step we compute the spin-up or down that results from the contraction or expansion, and consequent moment of inertia change, of the star in order to conserve angular momentum, and the spin-down that results from the loss of angular momentum through winds using the magnetic braking prescription from Section~\ref{sec:Model}, assuming solid-body rotation.  To do so we calculate the Rossby number using convective turnover times from the \textit{MIST} models and derive the magnetic complexity of each star using Equation~\ref{eq:n}, and its angular momentum loss rate using Equations~\ref{eq:J2} and \ref{eq:Q}. The new rotation period is then computed. 

Stars are born with a circumstellar disk which lasts for up to a few Myrs and is expected to prevent them from spinning up as they contract through ``disk locking'' \citep{Rebull.etal:02, Rebull.etal:04}. By $\sim 10$~Myrs all starts have lost their disks (which corresponds to the "birth line" of \citealt{Stahler:83} and \citealt{Palla.Stahler:90} at the end of the disk-locking time). 
%We have included this disk locking mechanism by assigning each star in our sample a random time between 1 and 10 Myrs to start their rotation evolution process.  Our model is not very sensitive to the exact duration of disk-locking.  This is expected from the fact that 10~Myrs is a very short timescale in which no visible changes arise in terms of OC rotation period distributions.  
%\textbf{We start our rotation evolution immediate post-disk phase in order to avoid introducing extra free parameters related to the disk-locking phase,  as pointed out by \cite{Rebull.etal:18}.  For that reason we use as our initial data the rotation periods in the 13 Myrs old cluster h Persei.  Our model is not very sensitive to initial conditions or the duration of disk locking.  We have tested our model using different initial distributions, for example the Orion Nebula Cluster as well as an homogeneous distribution of masses (0.3 - 1.6 $M_{\odot}$) and periods (0 - 15 days).  We find that initial conditions are mostly erased fairly quickly ($<$ 200 Myrs) and we recover the same bimodal distribution of rotation periods.  }
Our model is not very sensitive to the assumed initial conditions or the duration of disk locking.  For the results presented here, we start our rotation evolution in the immediate post-disk phase in order to avoid introducing extra free parameters related to disk-locking,  as pointed out by \cite{Rebull.etal:18}.  We use as our starting conditions the rotation periods of stars in the 13~Myr old cluster h Persei.  We have tested the  rotation evolution outcomes using different initial distributions, for example the rotation periods of stars in the Orion Nebula Cluster with disk locking, as well as an homogeneous distribution of masses (0.3 - 1.6 $M_{\odot}$) and periods (0 - 15 days).  We find that initial conditions are largely erased fairly quickly ($<$ 200 Myrs) and we recover the same bimodal distribution of rotation periods.
 The reason for this can be seen in Figure~\ref{fig:sun}. The initial period of each star will only determine how long that particular star will remain in the branch of fast rotators. Shorter disk-locking times will result in stars spinning up over a longer time interval and, therefore, should lead to more stars in the bottom branch at early ages ($\sim$100~Myrs).  But the general bimodal distribution of rotation periods will be, overall, unaffected. 

%The reason for it can be seen from Figure~\ref{fig:sun}. The initial period of each star will only determine how long that particular star will remain in the branch of fast rotators. Shorter disk-locking times will result in stars spinning up over a longer time interval and, therefore, should lead to more stars in the bottom branch at early ages ($\sim$100~Myrs).  But the general bimodal distribution of rotation periods will be, overall, unaffected. 

%using the \textit{MIST} isochrones evolutionary tables \cite{Dotter:16, Choi.etal:16, Paxton.etal:11, Paxton.etal:13, Paxton.etal:15} and  using  At each step we take the mass, radius, moment of inertia, convective turnover time, and effective temperature for each star from the corresponding \textit{MIST} isochrone table, we spin-up each star according to their change in radius in order to conserve angular momentum.  We calculate the Rossby number derive the magnetic complexity of each star, using Equation~\ref{eq:n}, and its  angular momentum loss rate using Equations~\ref{eq:J2} and \ref{eq:Q}. The new rotation period is then computed. 
  
%We include the evolution of the temperature of each star, %using an interpolation of \textit{Tevol} tables, 
%that is reflected on their color evolution.  
 
%We derive the relationship between $B-V$, $\tau$, and $M$ from \cite{Wright.etal:11}. 

We fix the normalization constant related to dipolar angular momentum losses to $c=1 \times 10^{41}$ gm~cm$^{2}$, which provides the best fit to the Skumanich branch in OCs observations. This  is consistent with a reasonable solar angular momentum loss rate $\dot{J}_{Dip}\sim  \times 10^{30}$ gm~cm$^{2}$~s$^{-2}$ \citep{Pognan.etal:18} as becomes clear from Figure~2.

For each star, there is an intrinsic maximum velocity at which centrifugal forces exceed self-gravity, called the break-up velocity, that is a function of $M$ and $R$, 
$$ \Omega_{break-up} = \sqrt{\frac{G M}{R^3}},$$ 
where $G$ is Newton's constant, and $M$ and $R$ are the star's mass and radius, respectively.
We assume stars are solid-body rotators and impose this limit in order to have a physically consistent model. This does not qualitatively affect our results.

We chose a sample of stars large enough to get a reliable probability density for each rotation period as a function of color and age. For visualization we use a smaller group of 600 stars, comparable to the number of stars in each observation.

%%%%%%%%%%%%%%%%%%%%%%%%%%%%%%%%%%%%%%%%%%%%%%%%%%%%%%%%%%%%%%%%%%%%%%%%%%%%%%%%%%%%%
% Results
%%%%%%%%%%%%%%%%%%%%%%%%%%%%%%%%%%%%%%%%%%%%%%%%%%%%%%%%%%%%%%%%%%%%%%%%%%%%%%%%%%%%%
\section{OPEN CLUSTER OBSERVATIONS}
\label{sec:OCO}

%\subsection{Observations}

We compare our predictions to rotation period observations for stars of different ages, ranging from 50~Myrs to 1~Gyrs. By doing so we can judge the performance of our model both in terms of reproducing the bimodal period distributions of the observations as well as their time evolution.  We use the rotation periods and ages from the following clusters available in the literature: Pleiades $\sim 70 -150$~Myrs \citep{Hartman.etal:10, Rebull.etal:16a},  Hyades $\sim 500 - 625$~Myrs \citep{Radick.etal:87, Delorme.etal:11, Douglas.etal:16},  Coma $\sim 500$~Myrs  \citep{Radick.etal:90},  M 34 $\sim 200 - 250$~Myrs \citep{Barnes:03, Meibom.etal:11a}, M 35 \citep{Meibom.etal:09} $\sim 100-150$~Myrs, M 37 $\sim 346 - 550$~Myrs \citep{Hartman.etal:09b, Wu.etal:09}, Praesepe $\sim 550 - 700$~Myrs  \citep{Delorme.etal:11, Douglas.etal:17, Rebull.etal:17}, and NGC 6811 $\sim 1$~Gyrs \citep{Meibom.etal:11b}.   We compare predicted and observed periods vs.\ $B-V$ color, converting $V-Ks$ colors to $B-V$ using the tables by \cite{Pecaut.etal:12} where necessary.

%IC 2391 \citep{Patten.Simon:96}, IC2602 \citep{Barnes.etal:99}, IC 4665 \citep{Allain.etal:96}, ALPHA PER \citep{Prosser.Grankin:13}, 
%NGC 3532 \citep{Barnes:98}; 

%\subsection{Complexity Evolution}
\section{RESULTS}
\label{sec:Results}

We have compared Skumanich-like rotation evolution with the one predicted by our model for each stellar mass (we show only the one solar mass case in Figure~\ref{fig:sun}).  
%The difference between the two models is that the latter accounts for the magnetic modulation of the angular momentum loss efficiency according to Equations~\ref{eq:Q} and \ref{eq:n}.  
In the former, the absence of the complexity-induced weakening of angular momentum loss in the T~Tauri phase means that wind-driven spin-down dominates over contraction-driven spin-up for the fastest initial rotators, such that those stars never achieve more rapid rotation than at their immediate post-disk phase: all converge to a rotation period of approximately 1 day at an age of 15~Myrs.  

In our model, stars that have shorter initial rotation periods remain fast rotating for longer, until they undergo a sharp spin down process at an age that depends on their mass and initial period (see Figure~\ref{fig:sun}).  The spread in the transition times is larger for lower mass stars, which explains why more massive stars transition first and why a bimodal distribution of rotation periods is observed for late spectral types in older clusters.  At constant ages between a few tens of Myrs and up to a few hundred Myrs, we can see concentration of stars at a long period and a short period (right panel of Figure~\ref{fig:sun}), and a few stars in between, which is the basis for bimodality.

\begin{figure*}
%\vspace{2in}
\center
\includegraphics[trim = 0.3in 2.3in  1.5in 2.5in,clip, width = 0.45\textwidth]{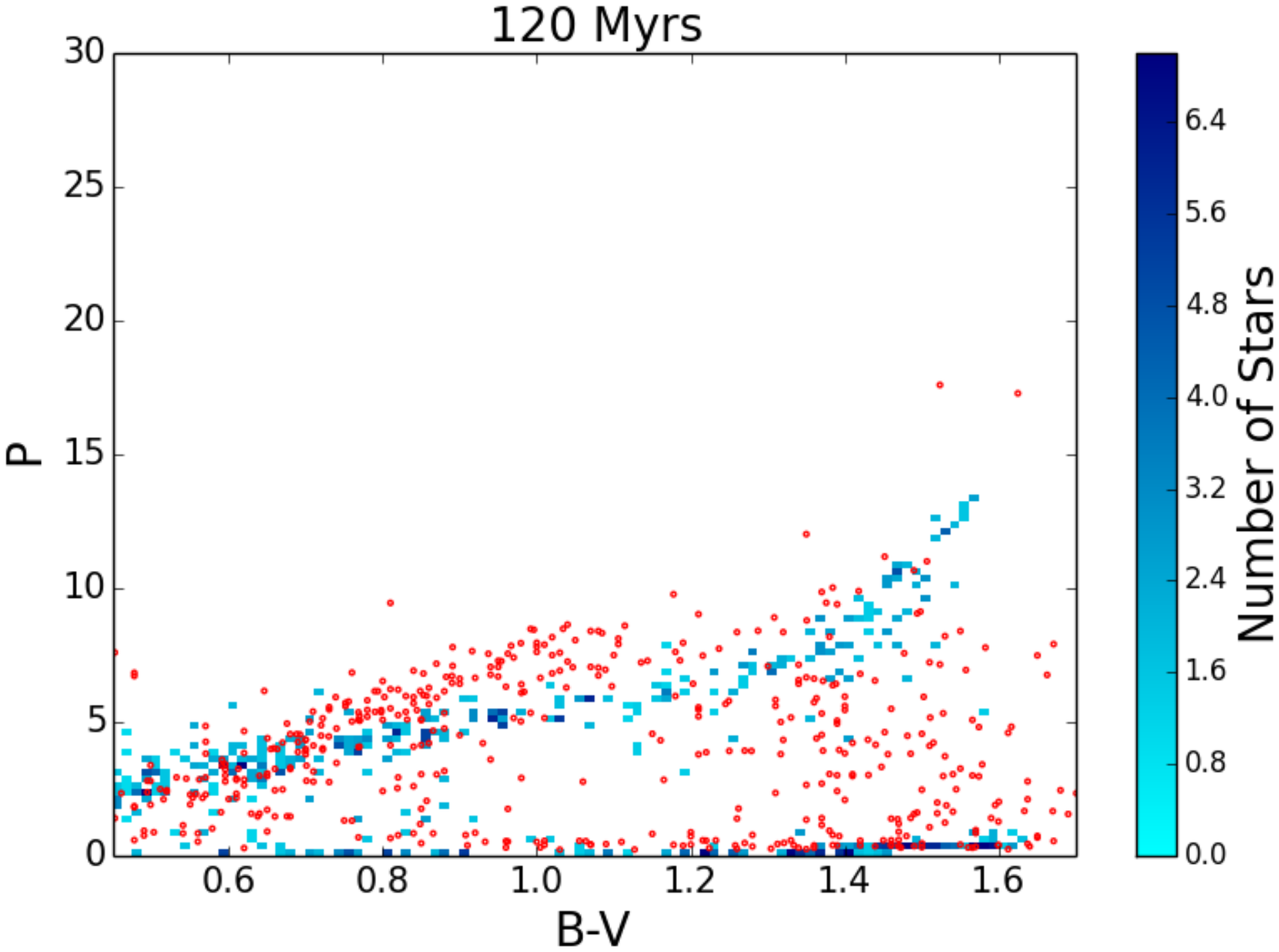}
\includegraphics[trim = 1.05in 2.3in 0.75in 2.5in,clip, width = 0.45\textwidth]{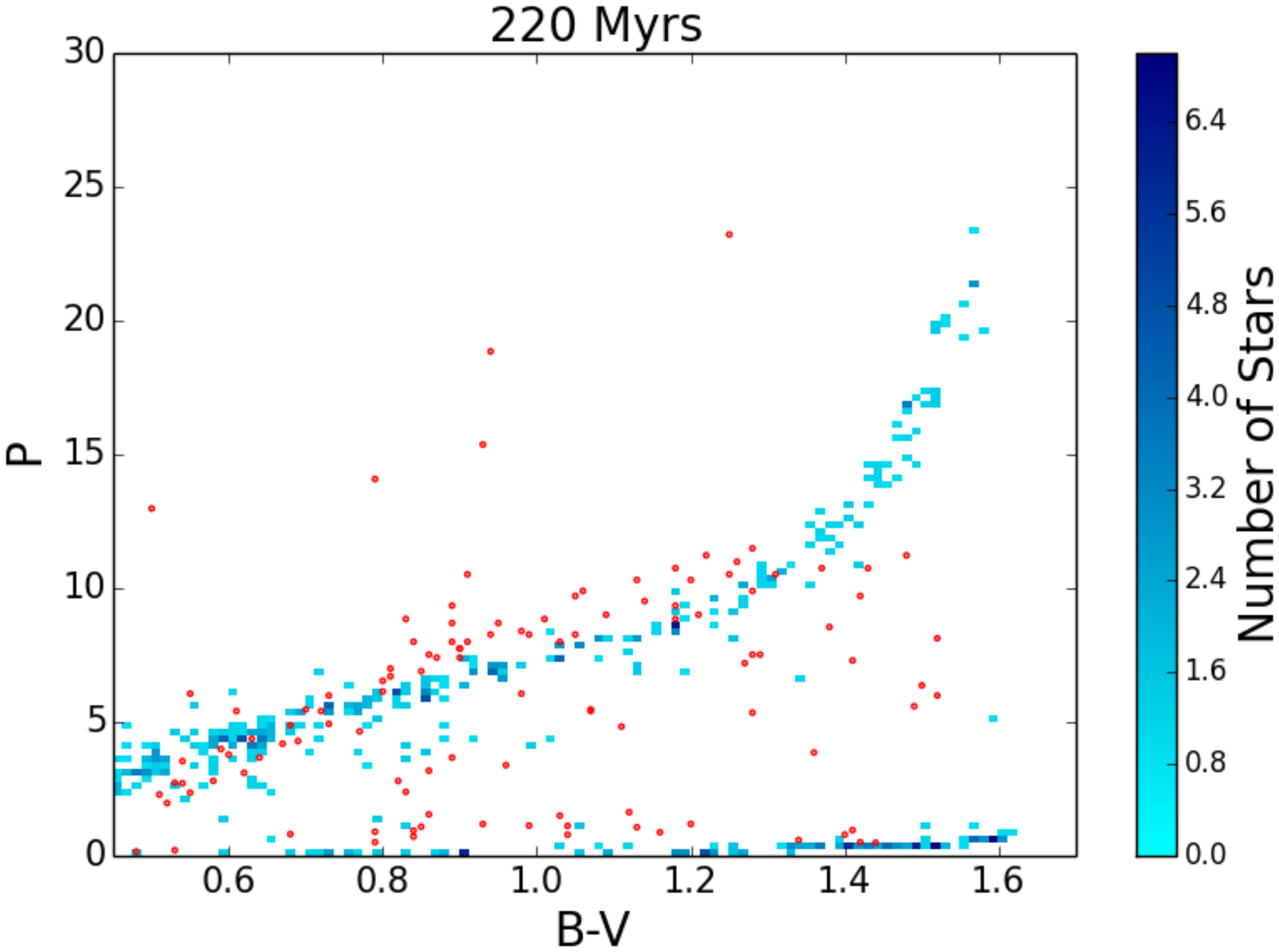}\\
\includegraphics[trim =  0.3in 2.3in  1.5in 2.5in,clip, width = 0.45\textwidth]{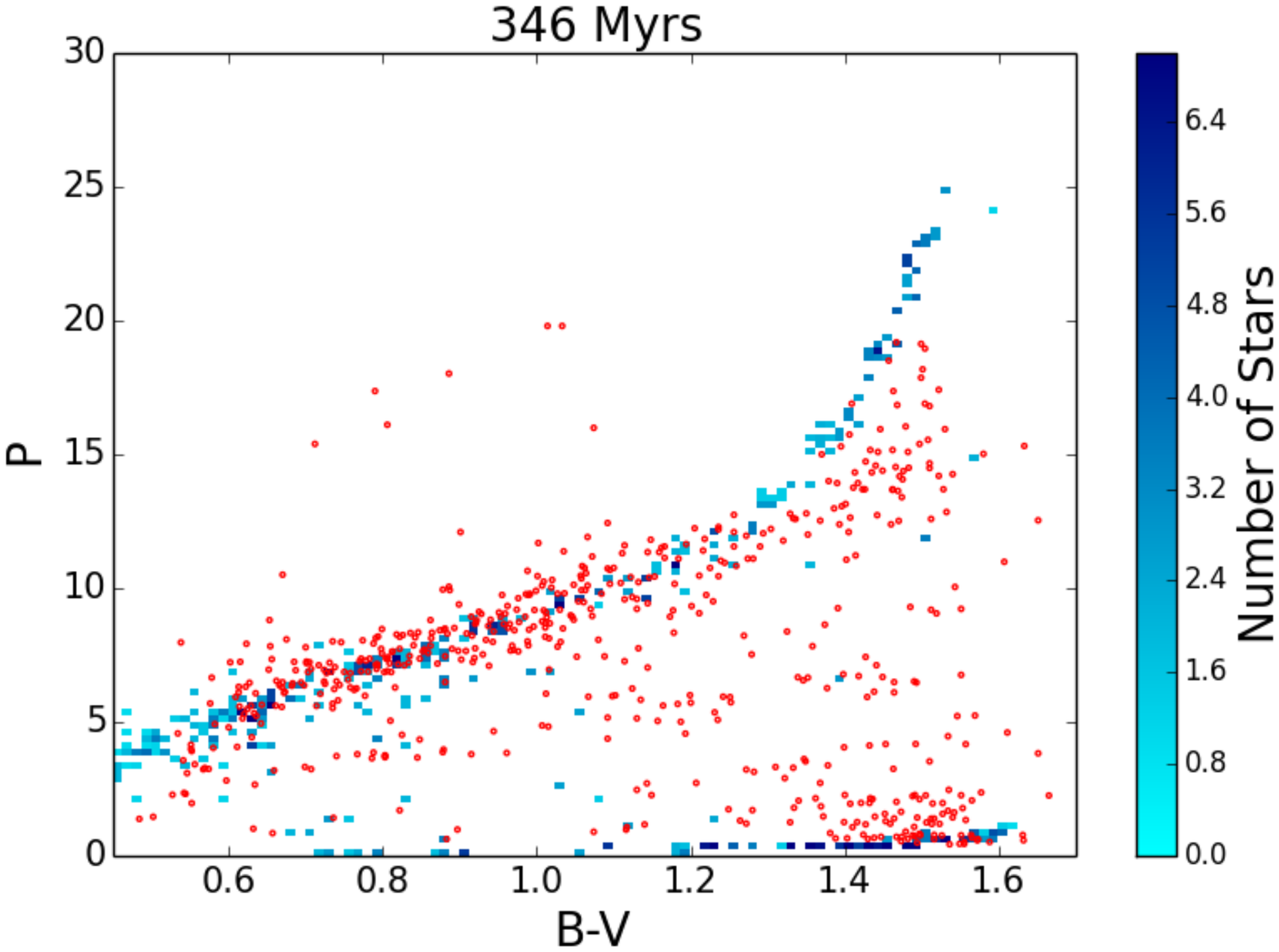}
\includegraphics[trim =  1.05in 2.3in 0.75in 2.5in,clip, width = 0.45\textwidth]{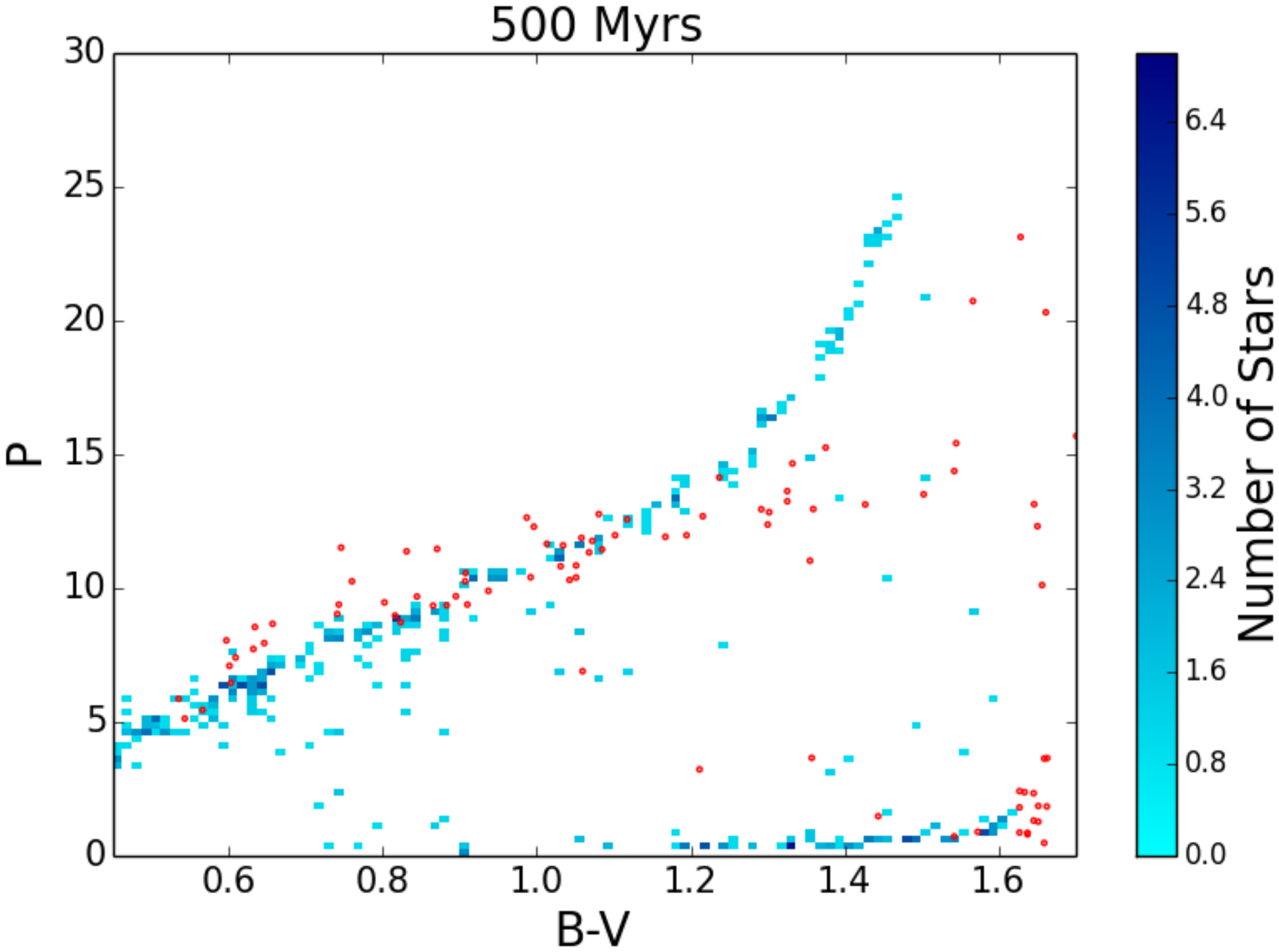}\\
\includegraphics[trim =  0.3in 2.3in  1.5in 2.5in,clip, width = 0.45\textwidth]{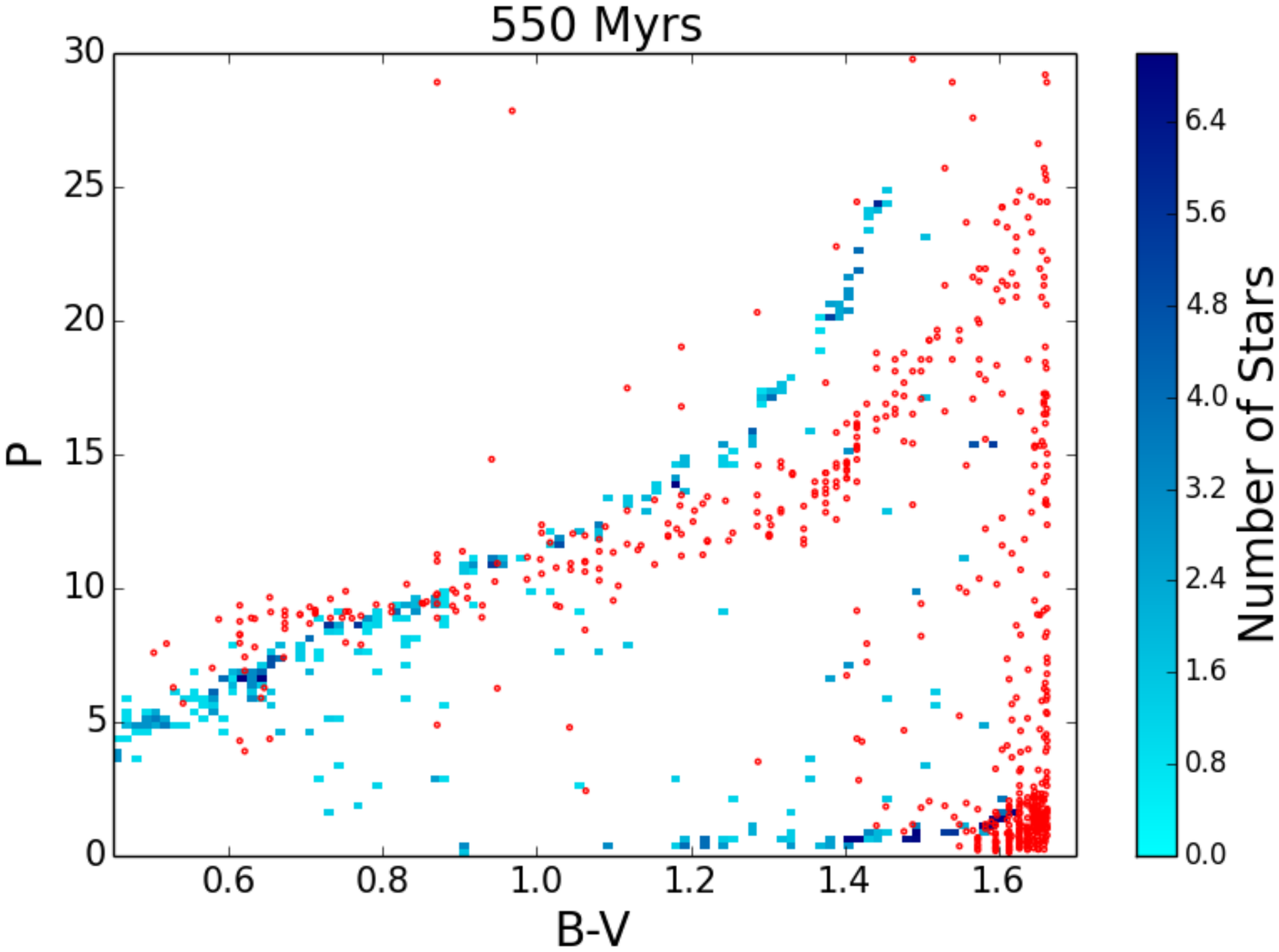}
\includegraphics[trim =  1.05in 2.3in 0.75in 2.5in,clip, width = 0.45\textwidth]{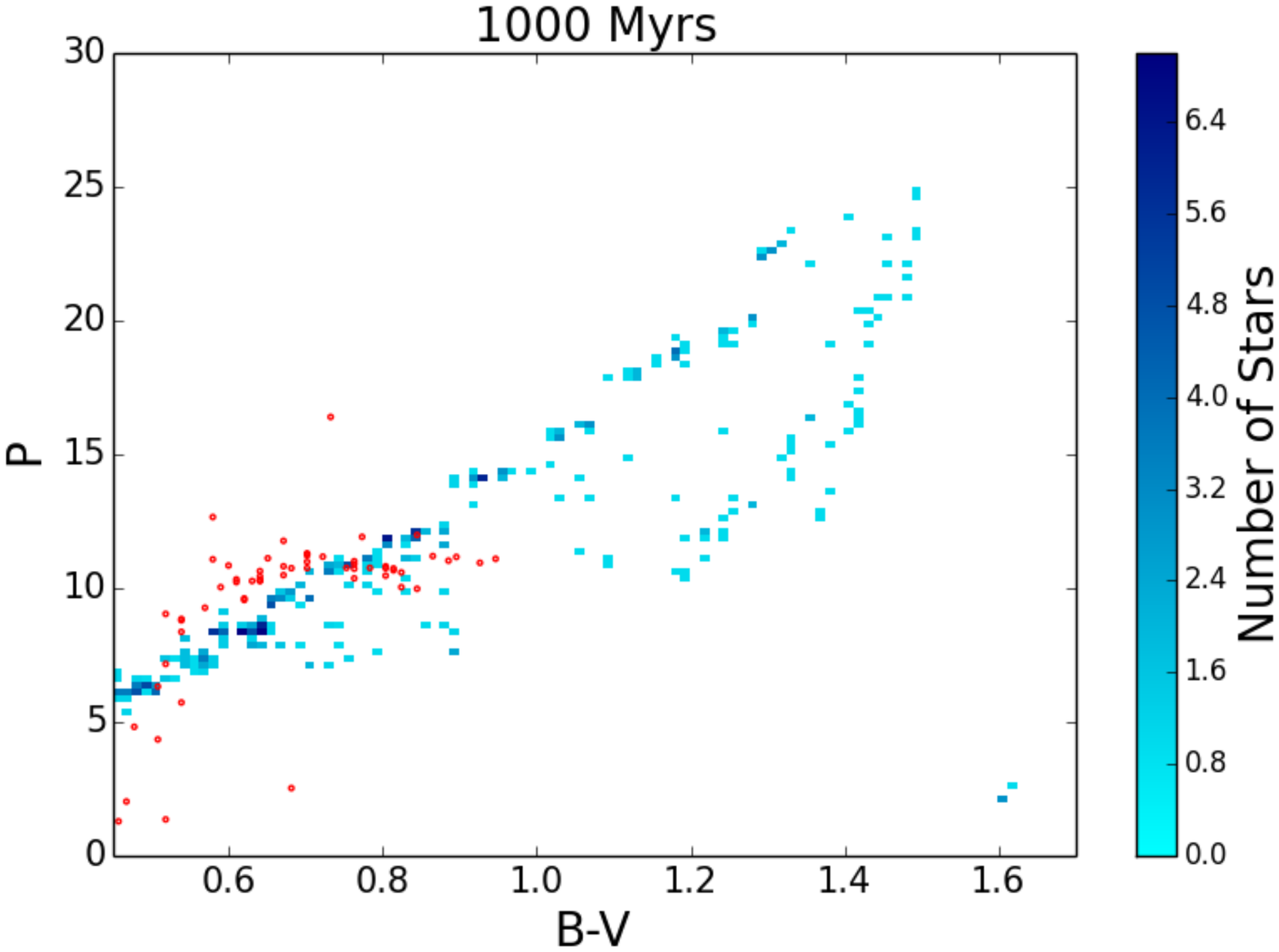}
\caption{Rotation periods as a function of {\it B-V} color observed in open clusters of increasing age (from top left to bottom right) are in red. In blue we show the probability density predicted by our model for each open cluster age (colorbar represents number of stars). We normalized our prediction to reflect the number of counts observed in each color bin.}
\label{fig:OC}
\end{figure*}

In Figure~\ref{fig:OC}, we compare population synthesis generated with our model (predicted density of stars in blue) to OC observations (observed stars in red).  For each age, we combine all available observations. We find, as expected from the solar mass star rotation evolution just discussed, that the bimodal distribution of rotation periods observed in OCs naturally arises (see Figure~\ref{fig:OC}), as do other characteristics of the rotation period distributions. The fast rotators branch near the bottom of each panel gets less populated with age, while the Skumanich top branch becomes more populated; stars with higher masses transition first.  In addition, we recover the mass dependence of the Skumanich branch.   

\section{DISCUSSION}
\label{sec:Discussion}

We find that including the magnetic modulation of angular momentum loss rates predicted from simulations (CG16) together with the complexity evolution observed by ZDI, results in the bimodal rotation morphology observed in OCs. By proposing a well-motivated complexity evolution function, we obtain good agreement for both branches, the gap between them, their mass dependency, and the time evolution of all the ingredients.

In this scenario, stars are magnetically more complex for Rossby numbers $< 0.1$.  Interestingly, this corresponds to the saturation regime  \citep[see, for example,][and references therein]{Wright.etal:11} in which X-ray emission reaches a plateau and faster rotation no longer results in stronger emission. 
%X-ray observations have shown a tight relation between x-ray luminosity fraction and Rossby number\citep[See, for example,][and references there in]{Wright.etal:11} in which faster rotation means stronger x-ray emission up to a point ($Ro \sim 0.1$) where the emission reaches a plateau called saturation.  
Our results suggest that stars are magnetically complex in the saturated regime and become simpler by the time they reach $Ro=0.1$, when they exit saturation and begin losing angular momentum efficiently.  In this picture, complexity is only determined by Rossby number. Higher mass stars with shorter convective turnover times, and consequently higher Rossby numbers for a given rotation period, transition sooner to the Skumanich branch. 

%Differential rotation is expected to take place after the saturated regime {\bf ??differential rotation should occur at shorter periods too - what is this discussion based on? ??} with the resulting magnetic cycles as a visible feature.  In our scenario, this is reflected as an increase of complexity for $Ro > 0.1$ (see Figure~\ref{fig:n}), responsible for the subtle branch in between fast and slow rotators for clusters of intermediate ages (see, for example, the 346 Myrs cluster {\bf ?? M37? ??} in Figure~\ref{fig:OC}).  {\bf ??hmm, I don't follow this??}

While the model presented here was originally motivated by the MDM model of \citet{Brown:14}, it has one important difference. Rotation evolution in the magnetic complexity driven spin-down model is entirely predictive from the moment of the alleviation of disk locking. At this time, its future rotation trajectory is determined entirely by its initial rotation period.  MDM includes an unavoidably stochastic component. While the physical mechanism responsible for the magnetic braking change is not specified, a discontinuous change in complexity that would cause a rapid change in rotation period would be a possibility.  In such a scenario,  stars within the gap between the rapid rotators and the Skumanich branch should all have similar (low) magnetic complexity.  Instead, the model presented here implies continuous evolution of the magnetic complexity at the stellar surface, determined by the star's Rossby number, that results in a smooth (although steep) decrease in rotation period.  The increasing number of ZDI observations should eventually help us distinguish between the two theories by showing whether or not a tight complexity/rotation relationship exists for these stars.

\subsection{Deviations from the Skumanich Law and Uncertainties in Cluster Ages}

Slight deviations from the Skumanich law have been discussed in the literature \cite[see, for example][and references therein]{Johnstone.etal:15b, Gallet.Bouvier:15}.  \cite{Brown:14}, for example, noted that no single function of color provides a good fit to all the OC observations and, therefore, he includes a correction to the Skumanich law that depends on the cluster's age. 
%We find that a later increase in complexity (after Ro$\sim 1 - 2$), consistent with \cite{Vansaders.etal:16}, improves the Skumanich branch time evolution when comparing it with observations and their reported ages. {\bf ?? shown? which observations? We do have problems matching the Skumanich branch for lower mass stars - B-V $< 1$ or so; we should mention that - is it the same problem Brown referred to? ??}  
We find the same problem here when assuming stars evolve as predicted by the Skumanich law. This can be seen from the discrepancy of the slow rotators branch (Figure~\ref{fig:OC}), especially for lower mas stars ({\it B-V} $> 1$). However, our main aim here is to understand the origin of the bimodal aspect of observed rotation periods and we therefore employ the standard Skumanich prescription and defer treatment of these additional details to future work.

There is some uncertainty in the reported ages of these clusters (see Section~\ref{sec:OCO}) which might lead to discrepancies with our model predictions.  Further discussion of these is beyond the scope of the present paper, though we note that errors in ages will mainly affect the time evolution of the Skumanich branch. We emphasize that the strength of the model presented here is that it qualitatively reproduces the morphology of the observations, provides a physics-based explanation, and predicts the existence of the fast rotators branch and the unpopulated gap in between the two branches. In terms of our model, a change in cluster ages could be accounted for in an overall shift of the apparent complexity function, $n$, with Rossby number and our conclusions would not be affected.

%However, this simplified scenario still provides the a reasonable fit and the best description so far for the slow rotators branch. On the other hand, there is some uncertainty on open clusters ages that makes it hard to draw conclusions on this.  

\subsection{Potential Complications from Binaries}

Binary systems represent a large fraction ($\sim 27\%$) of the stellar population and it is important to consider if their spin-down process is different than that of a single star.  While \cite{Bouvier.etal:97b} found no difference in rotation speeds between single and binary stars in the Pleiades ($\sim$ 120 Myrs), \cite{Patience.etal:02} reported that binaries with small separations (10-60 AU) rotate faster than wider binaries in $\alpha$ Persei ($\sim$ 60 Myrs). Later, \cite{Meibom.etal:07} found that binary stars rotate faster, on average, than single stars in M35 ($\sim$ 150 Myrs). 

Binary companions might affect spin-down rates of the system via tidal effects, magnetic interaction, or by altering the conditions of star formation or protoplanetary disk dispersal.  The gravitational and magnetic effects are only relevant for close binary systems and, while they can be significant in terms of spin-down rates \citep[e.g.][]{Zahn:77, Hut:81, Cohen.etal:12}, these systems make up only a small percentage of total cluster populations \citep[e.g.]{Meibom.etal:07} and we do not consider them further.  

A wider binary companion could still affect the rotation velocity of the system by truncating the disk that is thought responsible for preventing the star from spinning up at early stages through disk locking \citep{Artymowicz.Lubow:94, Armitage.Clarke:96, Lin.etal:93}. This would be expected to result in binary systems having, on average, shorter rotation periods at early ages and is consistent with observations of young clusters. Magnetic braking, however, introduces a self-regulating mechanism by which faster rotators spin-down faster and by a few Myrs this effect should be erased. Thus, in the absence of any other mechanisms, we should not expect wide-binaries to have a different rotation evolutionary path than single stars. Our model offers a new explanation for these observations.  If binary stars have, on average, shorter periods at early ages, that should result in them transitioning systematically later to the Skumanich branch (see Figure~\ref{fig:sun}).  
More observations of the rotation periods of binary systems with a range of different ages and masses would be useful to confirm this.  %If the finding is that they do rotate faster, then an extra mechanism than just initial conditions would be necessary to explain this behavior.  In addition, if the reason for the fast rotators branch to exist were that binary stars rotate faster, there would still need to be reason for them to quickly transition to the Skumanich branch.  We have not considered binary systems in this study and our model explains the bimodal regime without the need for a different spin-down track for binary systems. 

Lastly, metallicity might affect spin-down rates through the envelope opacity which determines the depth of the convection zone and, therefore, the convective turnover time $\tau$. In our model, the efficiency of the angular momentum mass loss rates is regulated by Rossby number and, thought it, $\tau$ has a role on determining the complexity of the stellar surface magnetic fields.  While this should be taken into account when modeling stellar spin-down in general, the OCs used here all have metallicities, Fe/H, within $\pm 0.16$ of the solar value 
\citep{Mermilliod.etal:97, Stauffer:97, Pinsonneault.etal:04}.  We neglect the effects of metallicity in our model for the present, though it might become more relevant in future work when extending rotation evolution models to older populations.   

%\cite{Vansaders.etal:16}. Is this later deviation from Gyrochronology a new increase on the complexity of the magnetic field. It is interesting to speculate on that very slow rotators might not have such a global dynamo and local effects could dominate their magnetism, leading to smaller scale fields. 
%The fact that there aren't many observed longer rotation periods than the one of the Sun (????? is this likely to be an observation bias? Kepler should be able to pick up longer rotation than this, right??????? )  suggests that the mechanism in effect at these later stages (larger Rossby numbers) is very efficient suppressing angular momentum loss, such as complexity is. It will be exciting to see what TESS contributes on this issue. 

\section{Conclusions}
\label{sec:Conclusions}

We have presented a spin-down model that, for the first time, accounts for the morphology of the stellar surface magnetic field.  Modulation of angular momentum loss according to the magnetic field complexity based on detailed MHD simulations has been included in the evolution of solid-body rotation. Spin-down that follows a Skumanich law can explain observed bimodal OC rotation periods by including a magnetic complexity that decreases with stellar rotation period, as indicated by surface magnetic field observations.  The model is entirely predictive from the moment a star loses its protoplanetary disk in the T~Tauri phase and its rotation evolution is governed by its change in moment of inertia and wind-driven angular momentum loss.

%%%%%%%%%%%%%%%%%%%%%%%%%%%%%%%%%%%%%%%%%%%%%%%%%%%%%%%%%%%%%%%%%%%%%%%%%%%%%%
% Acknowledgments
%%%%%%%%%%%%%%%%%%%%%%%%%%%%%%%%%%%%%%%%%%%%%%%%%%%%%%%%%%%%%%%%%%%%%%%%%%%%%%

\acknowledgments
%and Doug Burke
We thank the anonymous referee for very constructive comments. 
CG thanks Steven Saar, Soren Meibom, Stephanie Douglas, Ruth Angus, Scott Wolk, and Sean Matt for helpful comments and discussion. 
CG was supported by Chandra grants GO7-18017X and GO5-16021X.
JJD, VLK and DJB were supported by NASA contract NAS8-03060 to the {\it Chandra X-ray Center}. JDAG was supported by Chandra grants AR4-15000X and GO5-16021X. 
%, and thanks the Director, B.~Wilkes, and the CXC science team for advice and support.  
SPM and OC were supported by NASA Living with a Star grant number NNX16AC11G.
Numerical simulations were performed on the NASA HEC Pleiades system under award SMD-13-4526.

%%%%%%%%%%%%%%%%%%%%%%%%%%%%%%%%%%%%%%%%%%%%%%%%%%%%%%%%%%%%%%%%%%%%%%%%%%%%%%
% Bibliography
%%%%%%%%%%%%%%%%%%%%%%%%%%%%%%%%%%%%%%%%%%%%%%%%%%%%%%%%%%%%%%%%%%%%%%%%%%%%%%

%\bibliographystyle{aasjournal}
%\bibliography{AML}

\end{document}